\documentclass[conference, hidelinks]{IEEEtran}
\IEEEoverridecommandlockouts
\usepackage{cite}
\usepackage{amsmath,amssymb,amsfonts}
\usepackage{algorithmic}
\usepackage{graphicx}
\usepackage{textcomp}
\usepackage{xcolor}
\usepackage{hyperref}
\usepackage{array}
\usepackage{tabularx}
\def\BibTeX{{\rm B\kern-.05em{\sc i\kern-.025em b}\kern-.08em
    T\kern-.1667em\lower.7ex\hbox{E}\kern-.125emX}}

\usepackage[printwatermark]{xwatermark}
\usepackage{xcolor}
\usepackage{graphicx}
\usepackage{lipsum}

\newwatermark[allpages,color=red!20,angle=45,scale=3,xpos=0,ypos=0]{POSTPRINT}

\begin{document}

\title{An Automatic Deep Learning Approach for Trailer Generation through Large Language Models}

\author{
\IEEEauthorblockN{Roberto Balestri*\thanks{*Corresponding author: roberto.balestri2@unibo.it.}}
\IEEEauthorblockA{\textit{Department of the Arts} \\
\textit{University of Bologna}\\
Bologna, Italy \\
}
\and
\IEEEauthorblockN{Pasquale Cascarano}
\IEEEauthorblockA{\textit{Department of the Arts} \\
\textit{University of Bologna}\\
Bologna, Italy \\
}
\and
\IEEEauthorblockN{Mirko Degli Esposti}
\IEEEauthorblockA{\textit{Department of Physics and Astronomy} \\
\textit{University of Bologna}\\
Bologna, Italy \\
}
\and
\IEEEauthorblockN{Guglielmo Pescatore}
\IEEEauthorblockA{\textit{Department of the Arts} \\
\textit{University of Bologna}\\
Bologna, Italy \\
}
}

\maketitle

\textcolor{red}{Cite as: \\ R. Balestri, P. Cascarano, M. D. Esposti and G. Pescatore, "An Automatic Deep Learning Approach for Trailer Generation through Large Language Models," 2024 9th International Conference on Frontiers of Signal Processing (ICFSP), Paris, France, 2024, pp. 93-100, doi: 10.1109/ICFSP62546.2024.10785516. }
\\ \\

\begin{abstract}
Trailers are short promotional videos designed to provide audiences with a glimpse of a movie. The process of creating a trailer typically involves selecting key scenes, dialogues and action sequences from the main content and editing them together in a way that effectively conveys the tone, theme and overall appeal of the movie. This often includes adding music, sound effects, visual effects and text overlays to enhance the impact of the trailer.  In this paper, we present a framework exploiting a comprehensive multimodal strategy for automated trailer production. Also, a Large Language Model (LLM) is adopted across various stages of the trailer creation. First, it selects main key visual sequences that are relevant to the movie's core narrative. Then, it extracts the most appealing quotes from the movie, aligning them with the trailer's narrative. Additionally, the LLM assists in creating music backgrounds and voiceovers to enrich the audience's engagement, thus contributing to make a trailer not just a summary of the movie's content but a narrative experience in itself. Results show that our framework generates trailers that are more visually appealing to viewers compared to those produced by previous state-of-the-art competitors.
\end{abstract}


\section{Introduction \label{sec:1}}

Movie trailers stand as important elements, delivering brief, but profound, narratives that significantly influence a film's market reception and box office success.
The art of trailer creation is a synthesis of creativity and tactical planning.
These trailers are more than promotional snippets: they encapsulate the essence of the films they represent, using a blend of artistry and strategic insight to captivate audiences.
This creative process is intensive, demanding a deep dive into the film's narrative fabric, requiring teams to dissect and analyze scenes and dialogues to distill the film's core into a briefer narrative \cite{hesford2013action,hesford2013art}. 

Industry professionals employ time-consuming techniques to ensure these previews intrigue the viewers, encouraging them to explore the full movie \cite{finsterwalder2012effects,krebs2020effectful,marich2013marketing}. 
Creating a film trailer is a highly complex task, typically requiring several days of dedicated collaboration among a skilled team. 
This team examines the film's content and screenplay to fully grasp its narrative structure and themes. 
Hence, the challenge of automating this process is immense. However, it presents an exciting opportunity to enhance trailer production, aiming to replicate the human touch in terms of aesthetic quality, rhythmic precision, narrative coherence, and emotional impact.\cite{Irie2010,Hesham2018,argaw2024towards}.  

In the Artificial Intelligence (AI) era, generative models have revolutionized various creative industries, including many filmmaking tasks \cite{totlani2023evolution,zhu2023moviefactory}, such as script writing and story development \cite{chung2022talebrush}, character design and animation \cite{jaiswal2020towards,tang2024ai} and soundtrack composing \cite{kamath2024sound}. 
These technologies, particularly Large Language Models (LLMs) \cite{kasneci2023chatgpt}, are reshaping traditional approaches to content creation, offering unprecedented opportunities \cite{DEP2023}.
LLMs, such as Generative Pre-trained Transformer (GPT) models \cite{openaigpt23}, are at the forefront of natural language generation: they are trained on vast amounts of pre-existing textual data to produce contextually relevant (new) text. 
Historically, recurrent neural networks (RNNs) \cite{grossberg2013recurrent,loli2019recurrent} have been the main technologies for sequential data processing tasks and, in particular, for natural language processing tasks \cite{tarwani2017survey}. 
However, RNNs encountered several challenges, including vanishing gradients and difficulty in capturing long-range dependencies. 
As a result, LLMs became the preferred models by overcoming these limitations with transformer architectures and self-attention mechanisms \cite{wolf2020transformers,vaswani2017attention,galassi2020attention}, enabling them to understand and generate natural language with unprecedented accuracy and fluency.

So far, in the literature, there is a lack of usage of Generative AI driven systems that automatically generates movie trailers. For this reason, in this paper, we introduce a new framework for automated trailer creation. 
This framework utilize a LLM to orchestrate each phase of trailer production, from scene selection to the integration of on-screen and off-screen dialogues and music. The LLM's capabilities enable a full understanding of the movie plot, facilitating the assembly of a trailer that adheres to the film's narrative. 
Furthermore, to illustrate the capabilities of the technology, we have developed "hybrid" trailers that blend traditional voice-over techniques with the dynamic integration of actual movie dialogues.  
This work explores the capacity of the LLM-driven methodology to enhance automated systems with narrative insight and creativity. Our approach respects the tradition of trailer production (voice-over) while incorporating modern storytelling techniques (movie dialogues) \cite{GeorgeRichards}.

The flow of the manuscript is organized as follows: in Section \ref{sec:2} we provide a short overview of existing works in the field of trailer generation focusing on the usage of LLM-driven methods for video generation. Then, in Section \ref{sec:3} we report a detailed description of our framework. In Section \ref{sec:4} we carry on a detailed analysis of our framework's outcomes by comparing them with the ones obtained by two state-of-the-art methods in the field, namely Movie2trailer \cite{Rehusevych2020} and PPBVAM \cite{Xu2015}. Finally, in Section \ref{sec:5} we conclude the presentation of our paper.

The code for the framework is available at the GitHub repository \href{https://github.com/robertobalestri/Trailer-Generation-Framework}{https://github.com/robertobalestri/Trailer-Generation-Framework}. Due to copyright restrictions, direct access to the generated trailers for the majority of movies cannot be provided. However, a trailer for the public domain movie \textit{Night of the Living Dead} (1968) by George A. Romero can be viewed at \href{https://youtu.be/O9fS8s2LRqM}{https://youtu.be/O9fS8s2LRqM}.

\section{Related works \label{sec:2}}

The task of automatic creation of movie trailers has received less attention among the researchers if compared to the broader area of video content summarization. 

Video summarization aims to condense content offering tools to manage the growing volume of video data, typically used for educational purposes, surveillance footage, and general video archives, thus not trying to craft engaging narratives \cite{Brachmann2009}. 
This task has been largely investigated exploring different techniques \cite{meena2023review} such as learning-based paradigms \cite{apostolidis2021video,Mahasseni2017,Xie2023} and, very recently, focusing on the adoption of LLMs \cite{argaw2024scaling,zhang2024benchmarking}. 

Conversely, the automatic generation of movie trailers is an interdisciplinary research area spanning natural language processing, computer vision, and multimedia content creation aiming at the creation of an engaging and promotional snippets that attract viewers to watch the full video or movie. 
Prior works in this field have explored various techniques based on visual and/or auditory feature analysis. 
More precisely, in \cite{Xu2015} the authors introduce a surrogate measure of video attractiveness and develop a self-correcting point process-based model to describe video attractiveness dynamics, then they propose a graph-based algorithm to generate trailers. 
Similarly, in \cite{Hu2022} a graph convolutional neural network is used to extract visual and relational features of shots and selects the best ones for the trailer. 
A different paradigm is proposed in \cite{Smeaton2006} where the authors introduce an automated method for creating movie trailers by selecting shots based on audiovisual features using a support vector machine. 
Analogously, in \cite{Rehusevych2020} the authors make use of anomaly detection strategies for shot selection. 
In the literature, other approaches have been developed to generate the trailer based on emotion and content analysis. 
In \cite{Smith2017} an AI system is implemented integrating multi-modal semantics extraction to understand and encode emotional patterns specific to horror movies. The framework presented in \cite{Irie2010} employs affective content analysis to identify impactful speech and video segments, which are then assembled into a trailer using an algorithm designed to maximize the emotional impact. 
Finally, approaches based on narrative and contextual analysis have been adopted.
In \cite{Papalampidi2021} the authors leverage a graph-based representation of movies to identify narrative structures and predicting sentiment using screenplay text to enhance understanding of shot relationships. 
Similarly, in \cite{Hesham2018} natural language processing and machine learning are used to extract textual features from subtitles, to classify movies into genres, and to generate trailers accordingly.

In the field of video generation with the use of LLMs, there are notable innovations that, while not directly aimed at automatic trailer generation, highlight the potential of LLMs in the broader context of video content creation. 
For instance, Zhu et al. describe the MovieFactory framework that transforms textual descriptions into complete movies  in \cite{Zhu2023}. 
This system leverages GPT models to create detailed scripts, which are subsequently brought to life using video and audio generation techniques.
Similarly, in \cite{Long2024}, the authors introduce "VideoDrafter", a framework that utilizes LLMs to convert text prompts into multi-scene scripts. While these frameworks are not specifically designed for creating movie trailers, their methodologies highlight the versatility and power of LLMs in crafting engaging narrations. 

Most of the trailers generated by the aforementioned approaches lack of additional features like sound effects and music, being often a mere sequence of visual shots. 
Furthermore, the literature lacks of comprehensive approaches based on visual, emotional and narrative patterns all integrated in the process of generative movie trailers.
Unlike existing literature, the framework proposed in this paper advances trailer generation by incorporating a Large Language Model (LLM) to coordinate the various tasks involved in the trailer creation process. Our system ensures that each component—from scene selection to soundtrack composition—contributes to a trailer that captures the narrative essence of the film.

\begin{figure*}[htbp]
    \centering
    \includegraphics[width=0.8\linewidth]{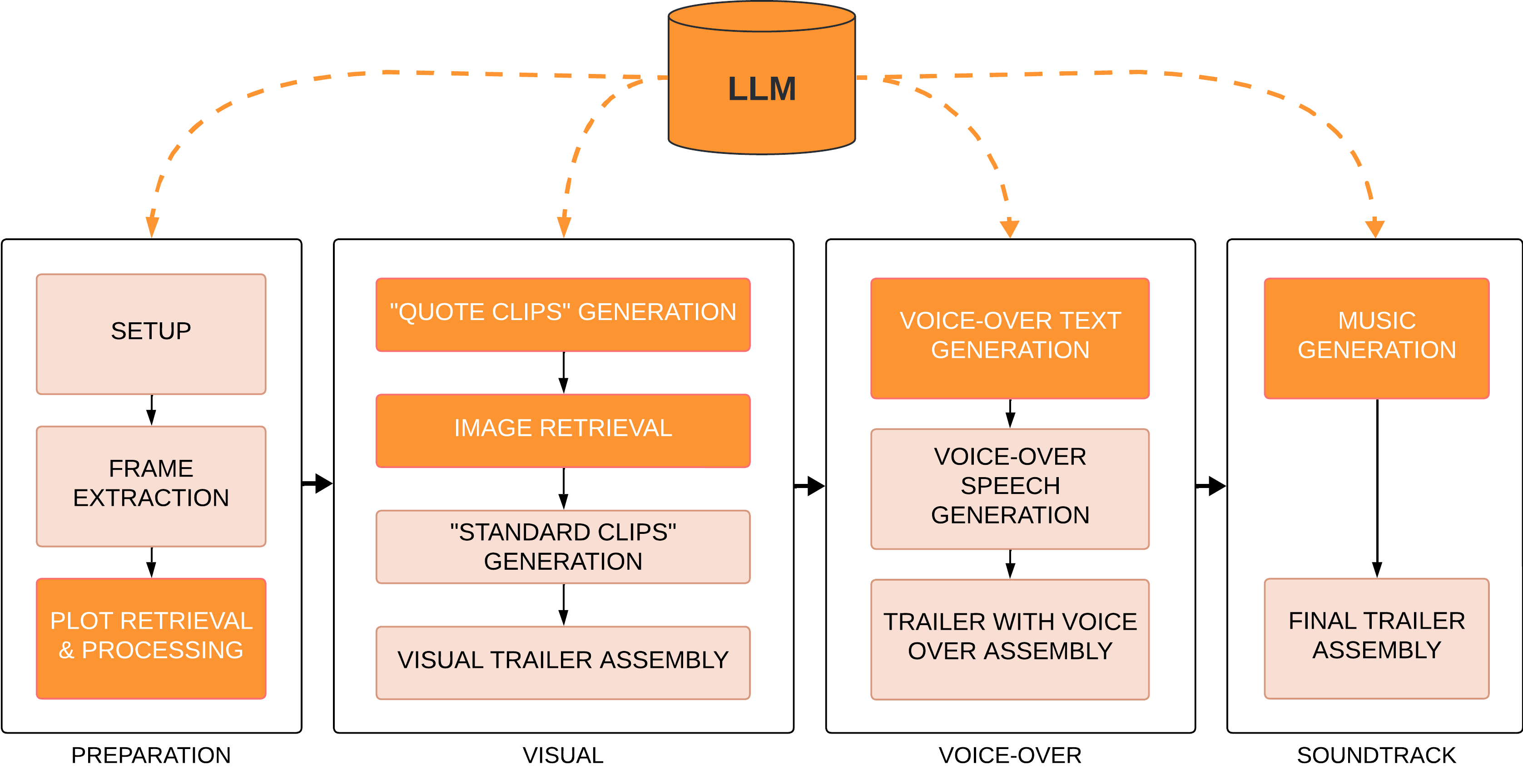}
    \caption{The framework's structure. We report all the four principal phases and the inner subphases.  The orange boxes highlight the specific subphases where the Large Language Model plays an active role.}
    \label{fig:1}
\end{figure*}

\section{The Proposed Framework \label{sec:3}}

This section presents the proposed framework. In Figure \ref{fig:1} we provide a conceptual map of its architecture, which is organized into four core stages, namely Preparation Stage, Visual Stage, Voice-Over Stage and Soundtrack Stage.  
We point out that the LLM we used is OpenAI's GPT-4 interfaced via API calls.
The framework's design is such that it could also be adapted to open-source models, suggesting that the approach is both innovative and accessible, with potential for broader application by other researchers. 
Hereafter, we describe the main phases and subphases of the process.

\subsection{Stage 1: Preparation.} This step involves an initial setup, extracting frames from the movie, and dividing the movie synopsis into scenes using the LLM. 

\begin{itemize}
    \item \textbf{Setup.} The initial setup involves setting up a few key configurations for movie trailer generation, including specifying Internet Movie DataBase (IMDB) \footnote{\href{https://www.imdb.com/}{IMBD website}} code, video file location, and project name.
    Movie information such as synopsis, relevant quotes, release date and director's name are scraped from IMDB using Cinemagoer \cite{Cinemagoer}, a Python library, and a custom script.

    \item \textbf{Frame Extraction.} This step focuses on extracting frames using FFmpeg \cite{FFmpeg}, a Python tool that ensures a balance between frame sampling and computational costs. The frames are extracted every nine seconds (an empirically chosen number). The first and last part of the movie are not computed in order to avoid opening and ending credits frames.

    \item  \textbf{Plot Retrieval \& Processing.} This step involves refining the movie's synopsis into a plot outline using the LLM. The synopsis is first filtered locally to comply with GPT's content filters \cite{achiam2023gpt}, then segmented into sub-plots by the LLM. Each sub-plot will represent a scene within the generated trailer, ensuring a coherent narrative arc. Sub-plots are organized for visual matching and trailer assembly. We asked the LLM to provide very visual subplots, so that we can retrieve the selected content from the movie more easily.
    For example, these are the first three sub-plots extracted for the movie Interstellar (2014):
    \begin{itemize} 
    \item[-]
    \textit{A dusty farm under a fading sky. }
    \item[-]
    \textit{A father, Cooper, checks over crops with a knowing frown. }
    \item[-]
    \textit{Young Murphy stares at mysterious patterns in the dust.}
    \end{itemize}
\end{itemize}

For the sake of readability we do not report the original input prompt given to the LLM which can be found in the Github repository. 
However, we asked the LLM to generate clear, simple descriptions of key visual scenes for a movie trailer, focusing on straightforward language to introduce main characters and locations without revealing the movie's conclusion, prioritizing simplicity and clarity, avoiding complexity and poetic language and producing descriptive phrases.

\subsection{Stage 2: Visual.} This step involves the creation of "Quote Clips," which are segments filled with impactful dialogues (selected by the LLM from the previously scraped quotes from IMDB movie's page) that capture the essence of the film's narrative, and "Standard Clips", which are the visual backbone of the story, guiding viewers without spoken words. These elements are then assembled into a coherent visual trailer.

\begin{itemize}
    \item \textbf{"Quote Clips" generation.} This step involves selecting key dialogues from the film to enrich the trailer's narrative depth. 
    After initial gathering, scraped movie's quotes undergo through cleaning and filtering, including speaker separation and content standardization. Length, structural completeness, and sentiment intensity (using TextBlob \cite{loria2018textblob} Python package) are checked to ensure relevance. The top 200 shortest quotes are selected, filtered from violent and sex-related words, and evaluated by the LLM for alignment with the film's themes and emotions. Audio extraction and transcription are done using StableWhisper \cite{Stablewhisper}, a sequence matching algorithm that aligns text quotes with audio segments, refined by Pyannote's \cite{Bredin2020} voice activity detection model. We refine the alignment with Pyannote because StableWhisper is not always precise in indicating the exact boundaries of the phrases. Video clips are extracted, and shot boundary detection (SBD) is applied \cite{PySceneDetect}. Based on the results from the SBD process, the system refines clips substituting any ”orphan shot” (shots that are too short to be visually appealing) at the start or at the end of the clip with a black screen to enhance visual continuity.

    For example, these are two among the selected quotes from the movie \textit{Interstellar}:

        \begin{itemize} 
            \item[-]
            \textit{Love is the one thing we're capable of perceiving that transcends time and space.}
            \item[-]
            \textit{We've always defined ourselves by the ability to overcome the impossible.}
        \end{itemize}

More precisely, We asked the LLM to analyze movie script phrases for trailer use, identifying impactful and memorable lines evaluating their emotional and thematic impact. We omit the full prompt for the sake of brevity.

    \item \textbf{Image retrieval.}  Inspired by a previous work from Dimitre Oliveira \cite{Dimitre}, the system aligns the film's narrative with visual representations by extracting keywords from sub-plot lines and using them as anchors. It employs the deep learning Clip-ViT-L-14 model  to embed textual and visual content into a shared semantic space, ensuring semantic congruence \cite{sentencetransformer, Reimers2019}. Frames exhibiting high semantic similarity to keywords (extracted from sub-plots) are selected based on temporal distribution. The selection process considers both the narrative progression and semantic alignment, quantified through cosine similarity measures. To ensure a representation of various parts of the movie, the distance in seconds between selected frames is at least 1.5\% of the total duration of the movie. EasyOCR \cite{ocr} and CRNN\cite{Shi2017} are used to ensure selected frames are free of superimposed text.

    The retrieved frame for the phrase "A dusty farm under a fading sky." from \textit{Interstellar} can be seen in Fig. \ref{fig:dusty_farm}. In this stage we asked the LLM to extract five key semantic keywords from movie plots, focusing on themes, characters, and significant events without redundancy.

    \begin{figure}[htbp]
        \centering
        \includegraphics[width=\linewidth]{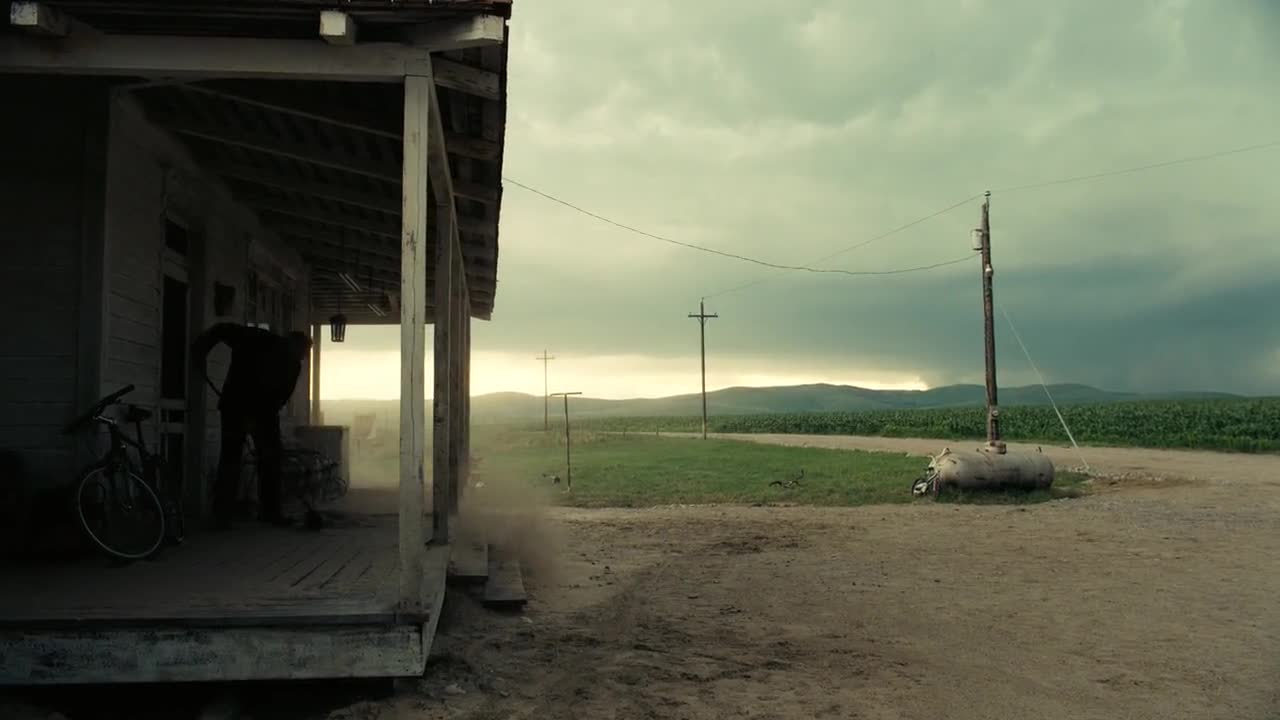}
        \caption{The frame retrieved for the sentence "A dusty farm under a fading sky." from the movie \textit{Interstellar}}.
        \label{fig:dusty_farm}
    \end{figure}

    \item \textbf{Standard Clip generation.} The system creates precise video segments by establishing a "buffered zone" around selected frames to capture the full context. Shot boundary detection algorithms \cite{PySceneDetect} are employed within this zone to identify start points for each clip, ensuring coherence and completeness. Like the Quote Clips generation, the system aims to prevent "orphan shots" in Standard Clips. Here, clip lengths are adjusted to improve narrative flow and ensure visual continuity.

    \item \textbf{Visual Trailer Assembly.} Here, the algorithm combines Standard and Quote Clips to create a narrative flow. Standard Clips are sorted to reflect their original narrative order, preventing inconsistencies. Quote Clips undergo audio source separation using a hybrid transformer model \cite{Rouard2022} to retain only the vocal part, then they are dispersed among standard clips using a systematic interval strategy. For example, if a trailer has 5 clips in total, of which 2 are Quote Clips (QC) and 3 are Standard Clips (SC), we have a trailer sequence like this: SC, QC, SC, QC, SC.\\
    Attention is paid to audio transitions, implementing fade-in and fade-out effects for smoother changes. All clips are concatenated into one unified video file, and a timestamp log is generated to record the start and end times of each Quote Clip in the trailer, aiding in later audio editing.

\end{itemize}

\subsection{Stage 3: Voice-Over.} This step involves the usage of the LLM to generate a voice-over script that complements the visual content. This script is converted to audio, and the voice-over is synchronized with the visual trailer.

\begin{itemize}
    \item \textbf{Voice-Over Text Generation.} In this phase, the LLM generates evocative phrases for the trailer's voice-over using the movie's plot summary, directorial credits, and release date. The generation ensure engagement and lyricism while avoiding spoilers. The quantity of phrases matches the trailer's length, integrating narrative and visual elements harmoniously.

For example, these are three generated voice-over phrases for \textit{Interstellar}:

        \begin{itemize} 
            \item[-]
            \textit{In the silence of space, hope whispers for a dying Earth.}
            \item[-]
            \textit{A ghost in the dust, a code to the stars under Christopher Nolan's vision.}
            \item[-]
            \textit{As hours become years, the journey for mankind's future unfolds in November.}
        \end{itemize}

 We asked to the LLM to craft captivating trailer phrases that mention the director's name and release month once, mimicking a sense of anticipation before the movie hits cinemas. The original prompt given as input to the LLM can be found in the Github repository.

    \item \textbf{Voice-Over Speech Generation.} The algorithm converts scripted text into audio using the Coqui xtts-v2 Text-to-Speech (TTS) open-source model \cite{Coqui, Zen2019}. Voice selection aligns with the movie's genre(s), employing a genre-to-voice mapping strategy for films with multiple genres. The TTS model generates audio files (that we call "Voice Clips") for each voice-over script segment.
    A generated voice-over phrase for the movie \textit{Interstellar} can be found at the link \href{https://freesound.org/people/bobe94/sounds/745217/}{https://freesound.org/people/bobe94/sounds/745217/}.

    \item \textbf{Trailer with Voice-Over Assembly.} In this phase, voice-over audio is integrated into the trailer. Timestamps indicating Quote Clip's placement are used to position Voice Clips avoiding any overlap between them. Average volume of Voice Clips informs volume adjustments for Quote Clips, ensuring seamless integration. Voice Clips are combined with the trailer's audio track and reattached to the video.
\end{itemize}

\subsection{Stage 4: Soundtrack.} This step involves the usage of a music generation model which composes a unique soundtrack that aligns with the film's themes thanks to the LLM’s musical direction. The final trailer is assembled by integrating this soundtrack with the voice-over and visual content, ensuring a balanced presentation. 

\begin{itemize}
    \item \textbf{Music Generation.} The LLM generates a detailed music description that aligns with the story's mood and theme. The system, employing the MusicGen text-to-audio model \cite{Copet2024}, synthesizes background music based on the LLM's description.
    More precisely, we asked the LLM to generate a music description inspired by the plot, focusing strictly on musical elements and instrument selection. The output of the LLM for the movie \textit{The Lord of the Rings: The Return of the King} is reported below: 

\begin{itemize}
    \item[] \textit{Instruments: Cello, dulcimer, low woodwinds, brass ensemble.}
    \item[] \textit{Key: D minor.} 
    \item[] \textit{Tempo: Moderate to slow.}
    \item[] \textit{Dynamics: Varied, with dynamic swells mirroring conflict.}
    \item[] \textit{Texture: Layered, introducing one instrument at a time.}
    \item[] \textit{Mood: Ominous with moments of somber reflection.}
    \item[] \textit{Atmosphere: Tense, foreboding.}
\end{itemize}

    An example of the soundtrack generated for the movie \textit{The Lord of the Rings: The Return of the King} can be found at \href{https://freesound.org/people/bobe94/sounds/750030/}{https://freesound.org/people/bobe94/sounds/750030/}.

    \item \textbf{Final Trailer Assembly.} In this phase, the system integrates the created music with the trailer's visuals and existing audio. An audio ducking algorithm lowers music volume during high trailer volume (e.g. during Quote Clips or Voice Clips) to preserve narrative clarity. The adjusted audio is combined with the video, including fade-in and fade-out effects, for a seamless viewing experience. This phase completes the trailer, ensuring the soundtrack enhances the narrative of the trailer.

\end{itemize}

\section{Evaluation and results \label{sec:4}}

In this section, we present a comparative analysis between our framework and other leading competitors in automatic trailer generation. The list of competing methods include: 

\begin{itemize}
    \item PPBVAM - Point Process-Based Visual Attractiveness Model \cite{Xu2015}
    \item Movie2trailer \cite{Rehusevych2020}.
    
\end{itemize}
We compare the trailers of \textit{The Wolverine} (2013), \textit{The Hobbit: The Desolation of Smaug} (2013), and \textit{300: Rise of an Empire} (2014) obtained by our system, with the outcomes produced by aforementioned approaches \footnote{\href{https://drive.google.com/drive/folders/16-v_L3KXYAowyKBU4mgA6mPT48AQTw6a?usp=sharing}{Competitors' generated trailers}}. For consistency, all trailers were standardized to a resolution of $480 \times 360$. \\
To ensure an objective evaluation, we adopted a rigorous approach. 
We engaged 16 volunteers (mainly students from a class of Cinema) with varied movie tastes to review the trailers, ensuring they had not seen any of the trailers before and were unaware of the order in which the methods were applied. 

\begin{table}[htbp]
\caption{Question items of the questionnaire submitted to the participants based on a Likert scale (1-7).}
\begin{center}
\begin{tabular}{|>{\centering\arraybackslash}m{2cm}|>{\centering\arraybackslash}m{3cm}|>{\centering\arraybackslash}m{2cm}|}
\hline
\textbf{Assessment} & \textbf{Question} & \textbf{Answer Type} \\
\hline
Appropriateness & How similar this trailer looks to an actual trailer? & Likert (1,7) \\
\hline
Attractiveness & How attractive is this trailer? & Likert (1,7) \\
\hline
Appropriateness & How likely you are going to watch the original movie after watching this trailer? & Likert (1,7) \\
\hline
\end{tabular}
\label{tab:questions}
\end{center}
\end{table}

Similarly to the evaluation carried out in \cite{Xu2015, Rehusevych2020}, the volunteers assessed each trailer on Appropriateness, Attractiveness, and Interest, which are metrics that reflect a trailer's ability to represent the movie, engage the audience, and pique interest.
The volunteers provided ratings on a Likert \cite{likert1932technique} scale from 1 (lowest) to 7 (highest) for each of the assessments, see Table \ref{tab:questions}. 

\begin{figure*}[htbp]
    \centering
    \includegraphics[width=0.7\linewidth]{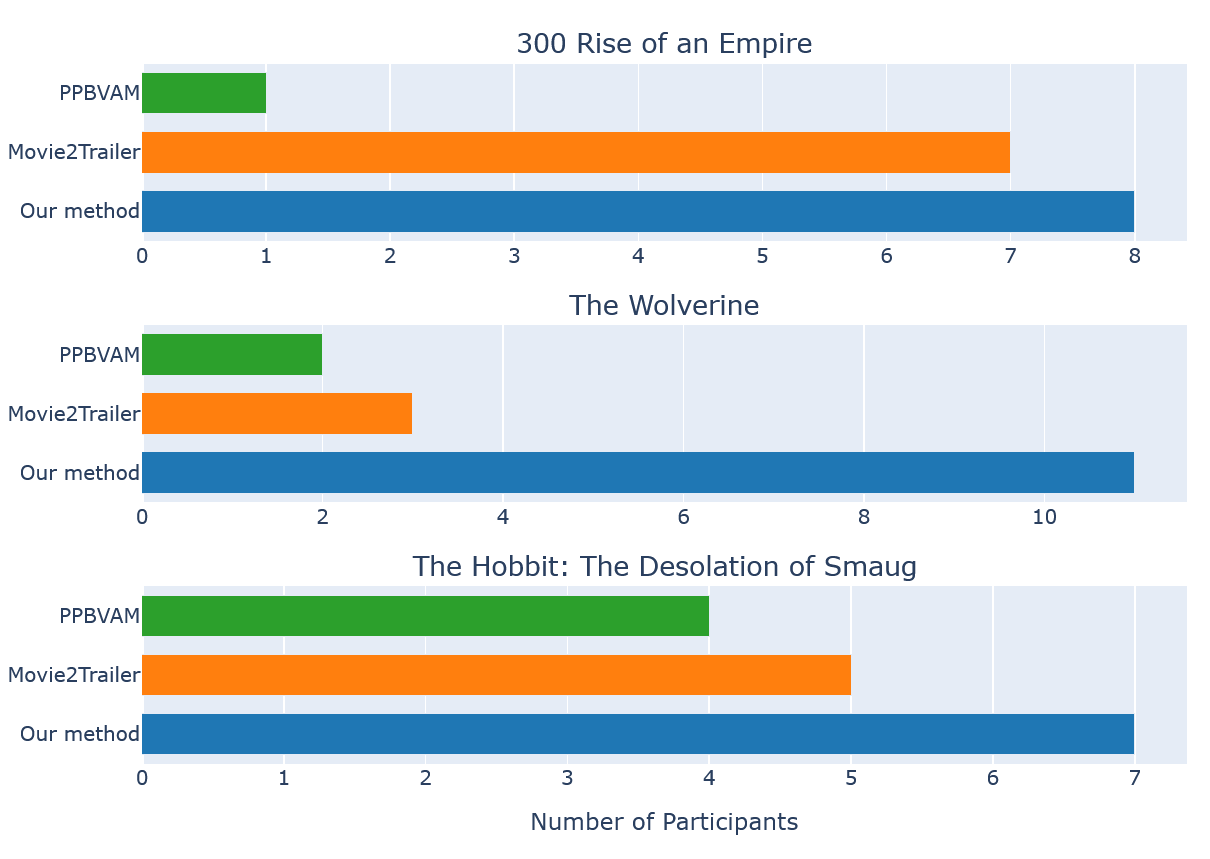}
    \caption{Comparison of the three automatic trailer generation methods based on the total scores achieved across three movies.}
    \label{fig:newplot}
\end{figure*}

We point out that, unlike PPBVAM and Movie2trailer, where the original movie soundtrack was used to replace the audio, our framework retains its uniqueness by incorporating AI generated voiceovers and soundtracks, providing a fully-automatic and comprehensive audio-visual approach.  
Furthermore, while we could have run our method multiple times to produce several trailers and then select the best one, the trailers generated are the outcome of a single run of the proposed software. This contrasts with our competitors, who may have chosen the best human-evaluated outputs from their frameworks.

As valuable metric, we considered the total score, namely the sum of the ratings across the three metrics. For each participant, we assumed that the method with the highest total score was considered the most effective. For example, if the trailer assessed 3 on Appropriateness, 3 on Attractiveness and 2 on Interest, its total score is 8.\\

In Figure \ref{fig:newplot} we depicted the number of participants who assigned to the competing methods the best score for considered movies. 
The barplot highlights which trailer generation method best meets the criteria set by viewers in a head-to-head comparison across different cinematic contexts.
Our system consistently outperformed its competitors, suggesting its superior ability to create appealing trailers. We think that the superior quality of the soundtrack and voiceovers in our framework's trailer for \textit{The Wolverine} explains its high ratings and significant outperformance compared to trailers generated for other movies.

Furthermore, we computed the average and median scores for each method across the evaluated metrics for the three movies, as shown in Tables \ref{tab:mean_scores} and \ref{tab:median_score}, respectively.
Our method demonstrated superior performance over PPBVAM in all categories and outperformed Movie2Trailer in Appropriateness and Interest. 
We think that the slight variance in Attractiveness is primarily due to the AI-generated soundtracks used by our system, which, despite being innovative, currently do not match the quality of the original human-composed soundtracks employed by competitors.
Apart from this, our framework's performance is commendable. 
The success in Appropriateness demonstrates its capacity to generate trailers that align closely, compared to the competitors, with conventional expectations of how a movie trailer should look and feel. Enhancing the AI's capability to produce soundtracks that are more akin to those created by humans could further improve its standings in Attractiveness, potentially making it a leader across all evaluated categories.

\begin{table}[htbp]
\caption{Average scores by method across the three categories.}
\centering
\begin{tabular}{lccc}
\hline
\textbf{Method} & \textbf{Appropriateness} & \textbf{Attractiveness} & \textbf{Interest} \\
\hline
Our Method & \textbf{3.56} & 2.96 & \textbf{2.90} \\
Movie2Trailer & 3.15 & \textbf{3.08} & 2.85 \\
PPBVAM & 2.81 & 2.88 & 2.62 \\
\hline
& \multicolumn{3}{c}{\textbf{Mean Scores}} \\
\hline
\end{tabular}
\label{tab:mean_scores}
\end{table}

\begin{table}[htbp]
\caption{Median scores by method across the three categories.}
\centering
\begin{tabular}{lccc}
\hline
\textbf{Method} & \textbf{Appropriateness} & \textbf{Attractiveness} & \textbf{Interest} \\
\hline
Our Method & \textbf{4.0} & \textbf{3.0} & \textbf{3.0} \\
Movie2Trailer & 3.0 & \textbf{3.0} & 2.0 \\
PPBVAM & 2.0 & \textbf{3.0} & 2.5 \\
\hline
& \multicolumn{3}{c}{\textbf{Median Scores}} \\
\hline
\end{tabular}

\label{tab:median_score}
\end{table}

\section{Limitations, Conclusions, and Future Work \label{sec:5}}

There are few limitations in our study which are mentioned below. First, the sample size we used for the survey is pretty small, which means our findings might not apply broadly. Future research should include larger and more diverse groups to get a clearer picture on the impact of the implemented framework.
Our main aim was to test if LLMs could generate trailers, not to beat the quality of human-made trailers. We went with an all-AI approach, which was innovative, but it resulted in Likert scale scores that didn't go much above average, though we did outperform existing competitors.
We had to show various trailers, the survey took a while to complete. We wanted to avoid boring the participants, a factor identified by \cite{meier_is_2024} as potentially biasing in studies. This made us design a survey that was a bit shallow in some areas. For instance, we didn't ask participants what elements they thought were missing in the trailers.
Moreover, after reviewing our trailers, we found several areas for improvements. In future developments we will improve coherence by integrating an action recognition model to avoid including scenes in the trailer where characters speak without corresponding audio in the trailers. Indeed, we observed that in our generated trailers, Standard Clips, which are intended to be video-only, sometimes feature  characters speaking, leading to disjointed and alienating results.
Future works will consider the usage of novel text-to-speech models for more expressive and natural voice-overs and advanced music generation models to better match the trailer's emotional tone. 
Finally, we will include more detailed quantitative and qualitative analyses to better assess our framework. Quantitatively, we'll measure how closely our generated trailers match real trailers for the same movies. Qualitatively, we'll identify specific elements our system is missing or could improve, based on a revamped survey.

In conclusion, while we're pleased with our results so far, having surpassed the state-of-the-art in automatic trailer generation, we know there's a long way to go before we can even hope to come close to the quality of human-crafted trailers.

\bibliographystyle{IEEEtran}
\bibliography{sample-ceur.bib}

\end{document}